\documentclass[showpacs,amsmath,amssymb,aps,showkeys,floatfix,prd,a4paper]{revtex4}
\usepackage{subfigure}
\usepackage{dcolumn}
\usepackage{bm}
\usepackage{epsfig}
\usepackage{amsfonts}
\usepackage{amssymb,amscd}
\def\lsim{\raise0.3ex\hbox{$<$\kern-0.75em\raise-1.1ex\hbox{$\sim$}}}
\def\gsim{\raise0.3ex\hbox{$>$\kern-0.75em\raise-1.1ex\hbox{$\sim$}}}

\newcommand{\be}{\begin{equation}}
\newcommand{\ee}{\end{equation}}

\def\beq{\begin{equation}}
\def\eeq{\end{equation}}
\def\beqa{\begin{eqnarray}}
\def\eeqa{\end{eqnarray}}

\newcommand{\ba}{\begin{eqnarray}}   
\newcommand{\eea}{\end{eqnarray}}

\def\gappeq{\mathrel{\rlap {\raise.5ex\hbox{$>$}}
{\lower.5ex\hbox{$\sim$}}}}
\def\lappeq{\mathrel{\rlap{\raise.5ex\hbox{$<$}}
{\lower.5ex\hbox{$\sim$}}}}
\def\Toprel#1\over#2{\mathrel{\mathop{#2}\limits^{#1}}}

\begin{document}
\begin{flushright}
\vskip1cm
\end{flushright}

\title{Probing the $X(4350)$  in $\gamma \gamma$ interactions at the LHC}
\author{V.P. Gon\c{c}alves}
\affiliation{Department of Physics and Astronomy, The University of Kansas, Lawrence, KS 66045, USA}
\affiliation{High and Medium Energy Group, Instituto de F\'{\i}sica e Matem\'atica,  Universidade Federal de Pelotas\\
Caixa Postal 354,  96010-900, Pelotas, RS, Brazil.}
\author{B.D.  Moreira}
\affiliation{High and Medium Energy Group, Instituto de F\'{\i}sica e Matem\'atica,  Universidade Federal de Pelotas\\
Caixa Postal 354,  96010-900, Pelotas, RS, Brazil.}

\begin{abstract}
The production of  $X(4350)$ in the  $\gamma \gamma$ interactions that occur  in proton-proton, proton-nucleus and nucleus-nucleus collisions at the CERN Large Hadron Collider (LHC)  is investigated and predictions for the kinematical ranges probed by the ALICE and LHCb Collaborations are presented. We focus on the $\gamma \gamma \rightarrow \phi J/\Psi$ process, which have been measured by the Belle Collaboration, and present parameter free predictions for the total cross sections at the LHC energies.
 Our results demonstrate that the experimental study of this process is feasible and can be used  to confirm or not the existence of the $X(4350)$ state. 
Finally, for completeness, we present predictions for the production of the $X(3915)$ state in the $\gamma \gamma \rightarrow \omega J/\Psi$ process and show that this exotic state can also be probed in $\gamma \gamma$ interactions at the LHC.  
 \end{abstract}



\date{\today}

\maketitle

%\vspace{1cm}

Over the last years the study of photon induced interactions at the LHC became a reality. The experimental results from the ALICE, ATLAS, CMS and LHCb Collaborations \cite{alice, alice2,lhcb,lhcb2,lhcb_ups,cms1,cms2,cms3,Atlas,exp5,exp6,exp7} are now being used to improve our understanding of the QCD dynamics as well to probe Beyond 
Standard Model Physics (For a recent review see Ref. \cite{review_forward}).
The basic ideas present in the description of photon induced interactions are very simple: ultra-relativistic charged hadrons are an intense source of photons and in a collision at large impact parameters ($b > R_{h_1} + R_{h_2}$, with $R_i$ being the hadron radius), denoted hereafter ultra - peripheral collisions (UPCs),  photon -- photon and photon -- hadron  interactions become dominant over the strong hadron -- hadron one. For photon - photon interactions, the total
cross section for a given process can be factorized in terms of the equivalent flux of photons of the incident hadrons  and  the  photon-photon production cross section \cite{uphic}. The experimental separation for such events is relatively easy, as photon emission is coherent over the  hadron and the photon is colorless we expect the events to be characterized by two intact recoiled hadrons (tagged hadrons) and the presence of two  rapidity gaps. 
 As pointed out in Refs. \cite{bertu,vicwer,vicmar,nosexotico}, photon -- induced interactions can also be used to study the production and properties of  exotic charmoniumlike states, which are a class of hadrons that decay to final states that contain a heavy quark and a heavy antiquark but cannot be easily accommodated in the remaining unfilled states in the $c\bar{c}$ level scheme (For reviews see e.g. Refs. \cite{bambrila,pr_navarra,navarra_mpla,karliner}). Our goal in this paper is to explore the possibility of producing the exotic $X(4350)$ in two-photon interactions in UPCs  with  
ultra-relativistic protons and nuclei. Such state have been observed by the Belle Collaboration \cite{belle} in the $\gamma \gamma \rightarrow \phi J/\Psi$ process, but none of the $\phi J/\Psi$ states observed by the LHCb Collaboration in the analysis of $B^+$ decays  is consistent with this state \cite{lhcb_exotic}. Therefore, the $X(4350)$ awaits confirmation \cite{karliner}. In what follows we use the values obtained by the Belle Collaboration for the product of the two-photon decay $\Gamma_{X(4350) \rightarrow \gamma \gamma}$ and branching fraction  ${\cal{B}}(X(4350) \rightarrow \phi J/\Psi)$
to derive parameter free predictions for the $X(4350)$ production in $\gamma \gamma$ interactions at the LHC. As we will show the resulting cross sections are large, which implies that a future experimental analysis is, in principle, feasible. Therefore, the study of this process can be used to confirm (or not) the existence and properties of this state. For completeness, we also demonstrate that a similar analysis can be performed for other exotic final states that have been observed in $\gamma \gamma$ processes by the Belle and BaBar Collaborations. In particular, we also will present predictions for the production of the $X(3915)$ state.

\begin{figure}[t]
\centering
\includegraphics[scale=0.65]{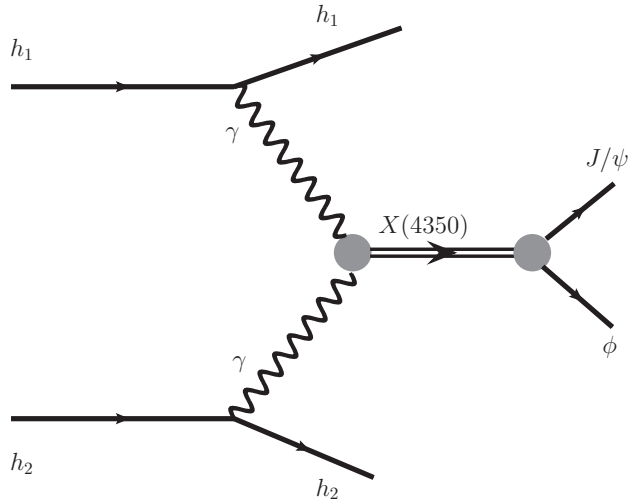}
\caption{Production of the exotic $X(4350)$ state in a $\gamma \gamma$ interaction at a hadronic collision. The $X(4350) \rightarrow \phi J/\Psi$ decay is also represented.}
\label{esquema_colisao}
\end{figure}

Initially, let's present a brief review of the formalism need to describe the photon -- induced interactions at hadronic colliders \cite{uphic}. In the particular case of the  $X(4350)$ production in $\gamma \gamma$ interactions at hadronic colliders, the cross section can be estimated using the equivalent photon approximation \cite{epa,uphic}. In this approximation the cross section for the production of the  exotic charmonium 
state, $X(4350)$,    in UPCs between two hadrons, $h_{1}$ and $h_{2}$, is given by (See e.g. \cite{uphic})
\begin{eqnarray}
\sigma \left( h_1 h_2 \rightarrow h_1 \otimes X(4350) \otimes h_2 ;s \right)   
&=& \int \hat{\sigma}\left(\gamma \gamma \rightarrow X(4350) ; 
W \right )  N\left(\omega_{1},{\mathbf b_{1}}  \right )
 N\left(\omega_{2},{\mathbf b_{2}}  \right ) S^2_{abs}({\mathbf b})  
 \mbox{d}^{2} {\mathbf b_{1}}
\mbox{d}^{2} {\mathbf b_{2}} 
\mbox{d} \omega_{1}
\mbox{d} \omega_{2} \,\,\, ,
\label{sec_hh}
\end{eqnarray}
where $\sqrt{s}$ is center-of-mass energy for the $h_1 h_2$ collision ($h_i$ = p,A), $\otimes$ characterizes a rapidity gap in the final state 
and $W = \sqrt{4 \omega_1 \omega_2}$ is the invariant mass of the $\gamma \gamma$ system. Moreover, 
 $N(\omega_i,b_i)$ is the equivalent photon spectrum generated by hadron (nucleus) $i$,
 which can be expressed as follows 
\begin{equation}
N(\omega_i,b) = \frac{Z^{2}\alpha_{em}}{\pi^2}\frac{1}{b^{2}\omega_i}
\left[ \int u^{2} J_{1}(u) F\left(\sqrt{\frac{\left( {b\omega_i}/{\gamma_L}\right)^{2} + u^{2}}{b^{2}}} \right )
\frac{1}{\left({b\omega_i}/{\gamma_L}\right)^{2} + u^{2}} \mbox{d}u\right]^{2}\, ,
\label{fluxo}
\end{equation}
where $\omega_{i}$ is the energy of the photon emitted by the hadron (nucleus) $h_{i}$ at an impact parameter, or distance, $b_{i}$ from $h_i$. 
Moreover, $\gamma_L$ is the Lorentz factor and $F$ is the nuclear form factor of the  equivalent photon source. The factor $S^2_{abs}({\mathbf b})$ is the absorption factor, given in what follows by \cite{BF90}
\begin{eqnarray}
S^2_{abs}({\mathbf b}) = \Theta\left(
\left|{\mathbf b}\right| - R_{h_1} - R_{h_2}
 \right )  = 
\Theta\left(
\left|{\mathbf b_{1}} - {\mathbf b_{2}}  \right| - R_{h_1} - R_{h_2}
 \right )  \,\,,
\label{abs}
\end{eqnarray}
where $R_{h_i}$ is the radius of the hadron $h_i$ ($i = 1,2$). In what follows we assume $R_p = 0.7$ fm and $R_{A} = 1.2 \, A^{1/3}$ fm.  
The presence of this factor in Eq. (\ref{sec_hh})  excludes the overlap between the colliding hadrons and allows to take into account only ultraperipheral collisions.
Finally, $ \hat{\sigma}_{\gamma \gamma \rightarrow X(4350)}(\omega_{1},\omega_{2})$ 
is the cross section for the production of a state $X(4350)$ from two real photons with energies $\omega_1$ and $\omega_2$. Using the Low formula \cite{Low}, the cross section for the production of  the $X(4350)$ 
state due to the two-photon fusion can be written in terms of the two-photon decay width $\Gamma_{X(4350) \rightarrow \gamma \gamma}$ as  follows
\begin{eqnarray}
 \hat{\sigma}_{\gamma \gamma \rightarrow X(4350)}(\omega_{1},\omega_{2}) = 
8\pi^{2} (2J+1) \frac{\Gamma_{X(4350) \rightarrow \gamma \gamma}}{M_{X}} 
\delta(4\omega_{1}\omega_{2} - M_{X}^{2}) \, ,
\label{Low_cs}
\end{eqnarray}
where $M_{X}$ and $J$ are, respectively, the mass and spin of the  produced  state.
Using that the photon energies $\omega_1$ and $\omega_2$  are related to   
$W$ and the rapidity $Y $ of the outgoing resonance $X(4350)$ by $
\omega_1 = \frac{W}{2} e^Y$ and $\omega_2 = \frac{W}{2} e^{-Y}$
the total cross section can be expressed by (For details see e.g. Ref. \cite{kluga})
\begin{eqnarray}
\sigma \left( h_1 h_2 \rightarrow h_1 \otimes X(4350) \otimes h_2 ;s \right)   
&=& \int \hat{\sigma}\left(\gamma \gamma \rightarrow X(4350) ; 
W \right )  N\left(\omega_{1},{\mathbf b_{1}}  \right )
 N\left(\omega_{2},{\mathbf b_{2}}  \right ) S^2_{abs}({\mathbf b})  
\frac{W}{2} \mbox{d}^{2} {\mathbf b_{1}}
\mbox{d}^{2} {\mathbf b_{2}} 
\mbox{d}W 
\mbox{d}Y \,\,\, .
\label{cross-sec-2}
\end{eqnarray}
 Such expression allows to easily estimate the rapidity distribution as well as to calculate the cross sections for the restricted range of rapidities considered by the ALICE and LHCb Collaborations.
In what follows  the cross sections will be estimated assuming  that the nucleus can be described by  a monopole form factor given by \cite{kluga}
\begin{equation}
F(q) = \frac{\Lambda^{2}}{\Lambda^{2} + q^{2}} \, ,
\label{ff_nuc}
\end{equation}
with $\Lambda = 0.088$ GeV. On the other hand, for proton projectiles, the form factor  will be assumed to be \cite{Goncalves:2015sfy,Goncalves:2016ybl}
\begin{eqnarray}
F(q) = 1/
\left[1 + q^{2}/(0.71\mbox{GeV}^{2}) \right ]^{2} \, .
\label{ff_pro}
\end{eqnarray}
A detailed discussion about the  theoretical uncertainty associated to the model used for $F$ is presented in Ref. \cite{nosexotico}.

\begin{table}[t] % aqui começa o ambiente tabela
\centering
\begin{tabular}{||c|c|c|c|c||} 
\hline 
\hline
Collision & Resonance & LHC&   LHCb  &  ALICE \\
\, & \,& Full rapidity range & $2 < Y < 4.5$ &  $-1 < Y < 1$ \\
\hline
\hline
$pp$ ($\sqrt{s} = $13 TeV) &  $X(4350)$, $0^{++}$   & (11.88 -- 29.50) fb & (2.47 -- 6.13) fb  & (2.67 -- 6.64) fb     \\  
 \,                        &   $X(4350)$, $2^{++}$  & (12.13 -- 33.09) fb & (2.52 -- 6.88) fb  & (2.73 -- 7.45) fb   \\ 
\hline
\hline
$pPb$ ($\sqrt{s} = $8.1 TeV) &   $X(4350)$, $0^{++}$ &  (36.98 -- 91.84) pb            & (10.20 -- 25.30) pb  &  (10.10 -- 25.00) pb \\  
  \,                         &   $X(4350)$, $2^{++}$ &  (37.76 -- 102.99) pb           & (10.30 -- 28.30) pb  &  (10.30 -- 28.00) pb   \\ 
\hline
$PbPb$ ($\sqrt{s} = $5.02 TeV) & $X(4350)$, $0^{++}$     & (93.40 -- 231.98) nb & (14.60 -- 36.20) nb & (34.60 -- 85.90) nb     \\  
  \,                           &    $X(4350)$, $2^{++}$  & (95.38 -- 260.14) nb & (14.90 -- 40.60) nb & (35.30 -- 96.30) nb    \\ 
\hline
\hline
\end{tabular}
\caption{Total cross sections for $X(4350)[J^P] \rightarrow \phi J/\psi$ production for different center - of - mass energies considering the  full LHC rapidity range as well as the rapidity ranges covered by the  ALICE and LHCb detectors.} % igual ao ambiente figura
\label{tab:x4350}
\end{table}

One have that the cross section is directly dependent of the values for  
the decay width $\Gamma_{X(4350) \rightarrow \gamma \gamma}$, mass and spin of the resonance. Such quantities can  be taken from experiment or can be theoretically estimated (See Refs. \cite{bertu,vicwer,nosexotico}). In our analysis we will focus on the case in that the resonance decays into a $\phi J/\Psi$ final state. As a consequence, the cross section will be proportional to $\Gamma_{X(4350) \rightarrow \gamma \gamma} \times {\cal{B}}(X(4350) \rightarrow \phi J/\Psi)$, where 
${\cal{B}}(X(4350) \rightarrow \phi J/\Psi)$ is the branching fraction. Such product have been measured by the Belle Collaboration to be \cite{belle} $\Gamma_{X(4350) \rightarrow \gamma \gamma} \times {\cal{B}}(X(4350) \rightarrow \phi J/\Psi) = [6.7^{+3.2}_{-2.4} \, \mbox{(stat)} \pm 1.1 \, \mbox{(syst)}]$ eV for $J^P = 0^+$, or 
$\Gamma_{X(4350) \rightarrow \gamma \gamma} \times {\cal{B}}(X(4350) \rightarrow \phi J/\Psi) = [1.5^{+0.7}_{-0.6} \, \mbox{(stat)} \pm 0.3 \, \mbox{(syst)}]$ eV for $J^P = 2^+$. In what follows we will use these values as input in our calculations. Therefore, our predictions for LHC energies will be  parameter free. A future experimental analysis of this process can be used to confirm (or not) the existence and properties of the $X(4350)$ state observed by the Belle Collaboration.

\begin{figure}
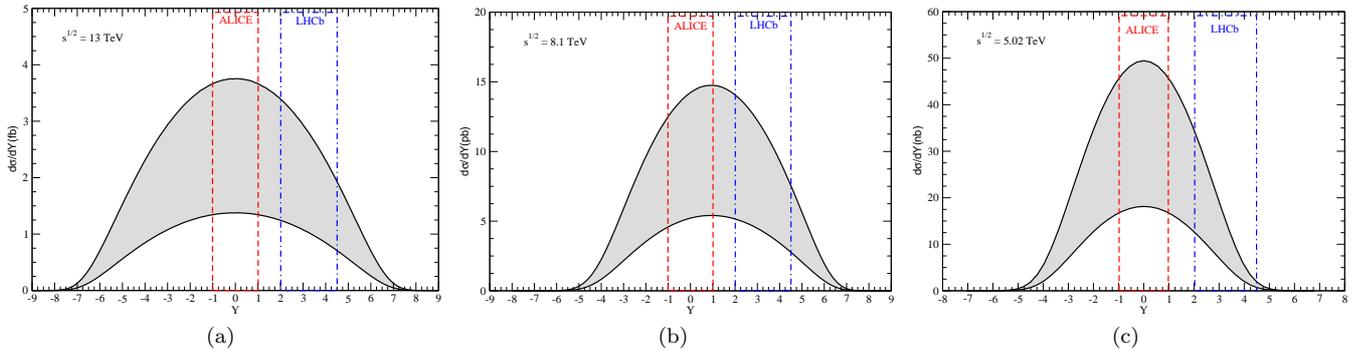

\begin{center}
\subfigure[ ]{\label{figa}
\includegraphics[width=0.32\textwidth]{pp_13000_dist_rap_X4350_2.eps}}
\subfigure[ ]{\label{figb}
\includegraphics[width=0.32\textwidth]{pPb_8100_dist_rap_X4350_2.eps}}
\subfigure[ ]{\label{figc}
\includegraphics[width=0.32\textwidth]{PbPb_5020_dist_rap_X4350_2.eps}}
\end{center}
\caption{
Rapidity distributions for the $X(4350)\,[2^{++}] \rightarrow \phi J/\psi$ production in (a) $pp \,(\sqrt{s} = 13 \,\mbox{TeV})$, (b) $pPb \, (\sqrt{s} = 8.1\,\mbox{TeV})$ 
and (c) $PbPb \, (\sqrt{s}=5.02\,\mbox{TeV})$ collisions at the LHC.
}
\label{fig:rap}
\end{figure}

In Table \ref{tab:x4350} we present our predictions for the total cross sections for the $X(4350) \rightarrow \phi J/\Psi$ production in $pp/pPb/PbPb$ collisions at the LHC energies considering the full LHC rapidity range as well as the rapidity ranges covered by the ALICE and LHCb detector. We consider the two possible values for $J^P$. The lower (upper) limit of our predictions are calculated using the minimum  (maximum) values for $\Gamma_{X(4350) \rightarrow \gamma \gamma} \times {\cal{B}}(X(4350) \rightarrow \phi J/\Psi)$ obtained from the Belle data \cite{belle} by taking the central value and subtracting (adding) the total uncertainty, calculated in quadrature from the statistical and systematical uncertainties. Due to the $Z^2$ dependence of the photon spectra, we have that the following hierarchy is approximately valid  for the $X(4350)$ production induced by $\gamma \gamma$ interactions: $\sigma_{PbPb} = Z^2 \cdot \sigma_{pPb} = Z^4 \cdot \sigma_{pp}$, with $Z = 82$.  In Fig. \ref{fig:rap} the corresponding rapidity distributions for the $X(4350)\,[2^{++}] \rightarrow \phi J/\psi$ production  are presented.  In the case of the $X(4350)\,[0^{++}]$ resonance, the distributions are similar but with a smaller normalization.  Due to the asymmetry in the proton and nuclear photon fluxes present in the initial state, we predict an asymmetric rapidity distribution in the case of $pPb$ collisions. In Fig. \ref{fig:rap} we also indicate the kinematical rapidity ranges probed by the ALICE ($-1 \le Y \le +1$) and LHCb ($+2 \le Y \le +4.5$) Collaborations.  The resulting predictions for the total cross sections in the ALICE and LHCb rapidity ranges are also  presented in Table \ref{tab:x4350}. In comparison with the results for the full LHC rapidity range, one have that the predictions are reduced by a factor between $2.7$ and $6.0$ depending on the initial state and the rapidity range covered by the detector, with the larger reduction being for $PbPb$ collisions at the LHCb.  Although this reduction is nonnegligible, the final values  are still large and imply a significant number of events if we consider that the expected integrated luminosity for the high luminosity run of the LHC is 50/fb (10/nb)  for $pp$ ($PbPb$) collisions. 
In particular, we predict  that the number of events per year in $pp$ ($PbPb$) collisions will be larger than 1850 (2300) at LHCb and 1970 (5200) at ALICE, which implies that the experimental analysis of this process is, in principle, feasible.

\begin{table}[t] % aqui começa o ambiente tabela
\centering
\begin{tabular}{||c|c|c|c|c||} 
\hline 
\hline
Collision & Resonance &  LHC & LHCb  &  ALICE \\
\, & \, & Full rapidity range & $2 < Y < 4.5$ &  $-1 < Y < 1$ \\
\hline
\hline
$pp$ ($\sqrt{s} = $13 TeV) &  $X(3915)$, $0^{++}$   & (177.89 -- 336.24) fb  & (36.80 -- 69.60) fb  & (39.40 -- 74.50) fb     \\  
 \,                        &   $X(3915)$, $2^{++}$  & (265.80 -- 492.74) fb & (55.00 -- 102.00) fb  & (58.90 -- 109.30) fb   \\ 
\hline
\hline
$pPb$ ($\sqrt{s} = $8.1 TeV) &   $X(3915)$, $0^{++}$ &  (561.84 -- 1061.98) pb & (150. 00 -- 290.00) pb  &  (150.00 -- 280.00) pb \\  
  \,                         &   $X(3915)$, $2^{++}$ &  (839.52 -- 1556.28) pb & (230.00 -- 420.00) pb  &  (220.00 -- 420.00) pb   \\ 
\hline
$PbPb$ ($\sqrt{s} = $5.02 TeV) & $X(3915)$, $0^{++}$     & (1453.61 -- 2747.60) nb &  (230.00 - 440.00) nb & (520.00 -- 990.00) nb     \\  
  \,                           &    $X(3915)$, $2^{++}$  & (2172.03 -- 4026.47) nb & (350.00 -- 650.00) nb & (790.00 -- 1460.00) nb    \\ 
\hline
\hline
\end{tabular}
\caption{Total cross sections for $X(3915)[J^P] \rightarrow \omega J/\psi$ production for different center - of - mass energies considering the  full LHC rapidity range as well as the rapidity ranges covered by the  ALICE and LHCb detectors.} % igual ao ambiente figura
\label{tab:x3915}
\end{table}

The analysis performed above can be directly extended for other exotic meson candidates. In particular, we can also provide predictions for the $X(3915)$ production. A candidate for such state was observed in $\gamma \gamma$ collisions by the Belle \cite{belle2} and BaBar \cite{babar} Collaborations in the $\gamma \gamma \rightarrow \omega J/\Psi$ process. Belle Collaboration obtained that \cite{belle2} 
$\Gamma_{X(3915) \rightarrow \gamma \gamma} \times {\cal{B}}(X(3915) \rightarrow \omega J/\Psi) = [61 {\pm 17} \, \mbox{(stat)} \pm 8 \, \mbox{(syst)}]$ eV for $J^P = 0^+$, or 
$\Gamma_{X(3915) \rightarrow \gamma \gamma} \times {\cal{B}}(X(3915) \rightarrow \omega J/\Psi) = [18 {\pm 5} \, \mbox{(stat)} \pm 2 \, \mbox{(syst)}]$ eV for $J^P = 2^+$.
Using these values as input in our calculations we can estimate the total cross sections for $X(3915)[J^P] \rightarrow \omega J/\psi$ production in $pp/pPb/PbPb$ collisions at the LHC energies. The predictions are presented in Table \ref{tab:x3915}. In comparison to the results for the $X(4350)$, the cross sections for the 
$X(3915)[J^P] \rightarrow \omega J/\psi$ production are one order of magnitude larger. The  cross sections are well within reach of present experiment detection techniques, considering the high luminosity expected for the next run of the LHC. 
Consequently, our results indicate that the analysis of this exotic meson also is, in principle, feasible at the LHC considering the photon induced interactions that occur in hadronic collisions at large impact parameter.

Finally, let's summarize our main results and conclusions. 
Over the last years the existence of exotic hadrons has been established  and a large number of candidate have been proposed. In particular, 
the exotic $X(4350)$ and $X(3915)$ mesons have been observed in $\gamma \gamma$ processes by the Belle Collaboration considering the production of the $\phi J/\Psi$ and $\omega J/\Psi$ final states, respectively.  However, such states have not observed at the LHC in the analysis of $B^+$ decays. Consequently, these states still awaits confirmation. In this study we have proposed the search of these resonances in the $\gamma \gamma$ interactions present in $pp/pPb/PbPb$ collisions  at the LHC. This is a clean process where the particles of the initial state are intact at the final state and can be detected at the forward direction as featured by the presence of two rapidity gaps between the 
projectiles and the produced resonance,  which is assumed to decay in a pair of vector mesons. Our results indicate that the experimental analysis of this process is, in principle, feasible at the LHC and that its study is ideal to confirm (or not) the existence and properties from these resonances.

\begin{acknowledgments}
One of the authors (VPG)  is partially supported by  the Brazil - U.S. Professorship given jointly by the Sociedade Brasileira de F\'{\i}sica (SBF) and the American Physical Society (APS).  This work was  partially financed by the Brazilian funding agencies CNPq, CAPES, FAPERGS and  INCT-FNA (process number 464898/2014-5).
 \end{acknowledgments}

\hspace{1.0cm}


\begin{thebibliography}{99}

\bibitem{alice} 
  B.~Abelev {\it et al.}  [ALICE Collaboration],
  %``Coherent $J/\psi$ photoproduction in ultra-peripheral Pb-Pb collisions at $\sqrt{s_{NN}} = 2.76$ TeV,''
  Phys.\ Lett.\ B {\bf 718}, 1273 (2013).


\bibitem{alice2} 
  E.~Abbas {\it et al.}  [ALICE Collaboration],
  %``Charmonium and $e^+e^-$ pair photoproduction at mid-rapidity in ultra-peripheral Pb-Pb collisions at $\sqrt{s_{NN}}$ = 2.76 TeV,''
  Eur.\ Phys.\ J.\ C {\bf 73}, 2617 (2013).
  
\bibitem{lhcb} 
  R. Aaij {\it et al.}  [LHCb Collaboration],
  %``Exclusive $J/\psi$ and $\psi(2S)$ production in $pp$ collisions at $\sqrt{s}=7$ TeV,''
  J.\ Phys.\ G {\bf 40}, 045001 (2013).


\bibitem{lhcb2} 
  R. Aaij {\it et al.}  [LHCb Collaboration],
  %``Exclusive $J/\psi$ and $\psi(2S)$ production in $pp$ collisions at $\sqrt{s}=7$ TeV,''
   J.\ Phys.\ G {\bf 41}, 055002 (2014)


\bibitem{lhcb_ups} 
  R.~Aaij {\it et al.} [LHCb Collaboration],
  %``Measurement of the exclusive Υ production cross-section in pp collisions at $ \sqrt{s}=7 $ TeV and 8 TeV,''
  JHEP {\bf 1509}, 084 (2015).



\bibitem{cms1} 
  S.~Chatrchyan {\it et al.}  [CMS Collaboration],
  %``Exclusive photon-photon production of muon pairs in proton-proton collisions at $\sqrt{s}=7$ TeV,''
  JHEP {\bf 01}, 052 (2012).
  
\bibitem{cms2} 
  S.~Chatrchyan {\it et al.}  [CMS Collaboration],
  %``Search for exclusive or semi-exclusive photon pair production and observation of exclusive and semi-exclusive electron pair production in $pp$ collisions at $\sqrt{s}=7$ TeV,''
  JHEP {\bf 11}, 080 (2012).  

\bibitem{cms3} 
  S.~Chatrchyan {\it et al.}  [CMS Collaboration],
  %``Study of exclusive two-photon production of $W^+W^-$ in $pp$ collisions at $\sqrt{s} = 7$ TeV and constraints on anomalous quartic gauge couplings,''
  JHEP {\bf 07}, 116 (2013).

\bibitem{Atlas} 
  G.~Aad {\it et al.} [ATLAS Collaboration],
  %``Measurement of exclusive $\gamma\gamma\rightarrow \ell^+\ell^-$ production in proton-proton collisions at $\sqrt{s} = 7$ TeV with the ATLAS detector,''
  Phys.\ Lett.\ B {\bf 749}, 242 (2015). 

\bibitem{exp5} 
  V.~Khachatryan {\it et al.} [CMS Collaboration],
  %``Evidence for exclusive $\gamma\gamma \to W^+ W^-$ production and constraints on anomalous quartic gauge couplings in $pp$ collisions at $ \sqrt{s}=7 $ and 8 TeV,''
  JHEP {\bf 1608}, 119 (2016).
  
\bibitem{exp6} 
  M.~Aaboud {\it et al.} [ATLAS Collaboration],
  %``Measurement of exclusive $\gamma\gamma\rightarrow W^+W^-$ production and search for exclusive Higgs boson production in $pp$ collisions at $\sqrt{s} = 8$ TeV using the ATLAS detector,''
  Phys.\ Rev.\ D {\bf 94}, no. 3, 032011 (2016).  
  
\bibitem{exp7} 
  M.~Aaboud {\it et al.} [ATLAS Collaboration],
  %``Measurement of the exclusive $\gamma \gamma \rightarrow \mu^+ \mu^-$ process in proton-proton collisions at $\sqrt{s}=13$ TeV with the ATLAS detector,''
  Phys.\ Lett.\ B {\bf 777}, 303 (2018).  

\bibitem{review_forward} 
  K.~Akiba {\it et al.} [LHC Forward Physics Working Group Collaboration],
  %``LHC Forward Physics,''
  J.\ Phys.\ G {\bf 43}, 110201 (2016).

\bibitem{uphic}
 G. Baur, K. Hencken, D. Trautmann, S. Sadovsky, Y. Kharlov, Phys.
Rep. {\bf 364}, 359 (2002); 
V.~P.~Goncalves and M.~V.~T.~Machado,
%``Parton Saturation Approach in Heavy Quark Production at High Energies,''
Mod. Phys. Lett. A {\bf 19}, 2525  (2004); 
 C.~A. Bertulani, S.~R.~Klein and J.~Nystrand, Ann. Rev. Nucl. Part. Sci. {\bf 55}, 
271 (2005);
 K.~Hencken {\it et al.},
  %``The Physics of Ultraperipheral Collisions at the LHC,''
  Phys.\ Rept.\  {\bf 458}, 1 (2008). 



\bibitem{bertu}     C.A. Bertulani, Phys.\ Rev.\ C {\bf 79}, 047901 (2009).	%24


\bibitem{vicwer}    V.~P.~Goncalves, D.~T.~Da Silva and W.~K.~Sauter, Phys.\ Rev.\ C {\bf 87}, 028201 (2013).	%22 

\bibitem{vicmar}    V.~P.~Goncalves and M.~L.~L.~da Silva,  Phys.\ Rev.\ D {\bf 89}, 114005 (2014).

\bibitem{nosexotico} 
  B.~D.~Moreira, C.~A.~Bertulani, V.~P.~Goncalves and F.~S.~Navarra,
  %``Production of exotic charmonium in $\gamma \gamma$ interactions at hadron colliders,''
  Phys.\ Rev.\ D {\bf 94}, no. 9, 094024 (2016)



\bibitem{bambrila} 
  N.~Brambilla, S.~Eidelman, B.~K.~Heltsley, R.~Vogt, G.~T.~Bodwin, E.~Eichten, A.~D.~Frawley and A.~B.~Meyer {\it et al.},
  %``Heavy quarkonium: progress, puzzles, and opportunities,''
  Eur.\ Phys.\ J.\ C {\bf 71}, 1534 (2011)
  
\bibitem{pr_navarra} 
  M.~Nielsen, F.~S.~Navarra and S.~H.~Lee,
  %``New Charmonium States in QCD Sum Rules: A Concise Review,''
  Phys.\ Rep.\  {\bf 497}, 41 (2010)

\bibitem{navarra_mpla} M. Nielsen and F. S. Navarra, Mod. Phys. Lett. A{\bf 29}, 1430005 (2014).



\bibitem{karliner} 
  M.~Karliner, J.~L.~Rosner and T.~Skwarnicki,
  %``Multiquark States,''
  Ann.\ Rev.\ Nucl.\ Part.\ Sci.\  {\bf 68}, 17 (2018)

\bibitem{belle} 
  C.~P.~Shen {\it et al.} [Belle Collaboration],
  %``Evidence for a new resonance and search for the Y(4140) in the gamma gamma ---> phi J/psi process,''
  Phys.\ Rev.\ Lett.\  {\bf 104}, 112004 (2010)

\bibitem{lhcb_exotic} 
  R.~Aaij {\it et al.} [LHCb Collaboration],
  %``Observation of $J/\psi\phi$ structures consistent with exotic states from amplitude analysis of $B^+\to J/\psi \phi K^+$ decays,''
  Phys.\ Rev.\ Lett.\  {\bf 118}, no. 2, 022003 (2017)


\bibitem{epa} 
  V.~M.~Budnev, I.~F.~Ginzburg, G.~V.~Meledin and V.~G.~Serbo,
  %``The Two photon particle production mechanism. Physical problems. Applications. Equivalent photon approximation,''
  Phys.\ Rept.\  {\bf 15}, 181 (1975).


\bibitem{BF90} G. Baur, L.G. Ferreira Filho, Nucl. Phys. A {\bf 518}, 786 (1990). 	%27


\bibitem{Low}       F.~E.~Low,  Phys.\ Rev.\  {\bf 120}, 582 (1960). 				% 26


\bibitem{kluga}     M.~Klusek-Gawenda and A.~Szczurek, Phys.\ Rev.\ C {\bf 82}, 014904 (2010).	%10               


\bibitem{Goncalves:2015sfy}   V.~P.~Goncalves, B.~D.~Moreira and F.~S.~Navarra,  Eur.\ Phys.\ J.\ C {\bf 76}, 103 (2016). 		%28

\bibitem{Goncalves:2016ybl}   V.~P.~Goncalves, B.~D.~Moreira and F.~S.~Navarra,  Eur.\ Phys.\ J.\ C {\bf 76}, 388 (2016).	% 29

\bibitem{belle2} 
  S.~Uehara {\it et al.} [Belle Collaboration],
  %``Observation of a charmonium-like enhancement in the gamma gamma ---> omega J/psi process,''
  Phys.\ Rev.\ Lett.\  {\bf 104}, 092001 (2010)

\bibitem{babar} 
  J.~P.~Lees {\it et al.} [BaBar Collaboration],
  %``Study of $X(3915) \to J/\psi \omega$ in two-photon collisions,''
  Phys.\ Rev.\ D {\bf 86}, 072002 (2012)  
  

\end{thebibliography}
\end{document}